\documentclass{article}

\usepackage[square,comma,numbers]{natbib}
\usepackage{gensymb}
\usepackage{amsmath}
\usepackage{authblk}
\usepackage{lscape}
\usepackage{blindtext}
\usepackage{makecell}
\usepackage{xcolor}
\usepackage{graphicx}
\usepackage{amssymb}
\usepackage{float}
\usepackage{url}
\usepackage{afterpage}
\usepackage{booktabs,multirow,array,caption}
\usepackage{textgreek}
\usepackage{multirow}
\usepackage{enumitem}
\usepackage{xcolor}
\usepackage{longtable}
\usepackage{lipsum}
\usepackage{capt-of}
\usepackage{textcomp}
\usepackage{algorithm}
\usepackage{algpseudocode}
\usepackage{tabularx}
\usepackage[
  colorlinks=true,
  urlcolor=blue,
  citecolor=blue,
  linkcolor=blue
]{hyperref}
\usepackage[a4paper,margin=1.1in,footskip=0.25in]{geometry}

\usepackage{mathpazo}   
\usepackage{abstract}

\usepackage{caption}
\usepackage{booktabs}
\usepackage{caption}
\usepackage{float}
\usepackage{titlesec}
\usepackage{capt-of}

\usepackage{array}
\usepackage{arydshln}
\title{\vspace{-2em}\bf%
The embodied brain: Bridging the brain, body, and behavior with biorealistic neuromechanical models
}
\date{}

\author[a,*]{Sibo~Wang-Chen}
\author[a,*]{Pavan~Ramdya}

\affil[a\ ]{Neuroengineering Laboratory, Brain Mind Institute \& Institute of Bioengineering, EPFL, Lausanne, Switzerland}
\affil[*]{Correspondence: \href{mailto:sibo.wang@epfl.ch}{sibo.wang@epfl.ch}, \href{mailto:pavan.ramdya@epfl.ch}{pavan.ramdya@epfl.ch}}

\begin{document}
\maketitle
\bibliographystyle{naturemag-x}  

\subsection*{Graphical Abstract}
\begin{figure}[!h]
    \centering
    \hspace*{-0.06\textwidth}
    \includegraphics[width=1.12\linewidth]{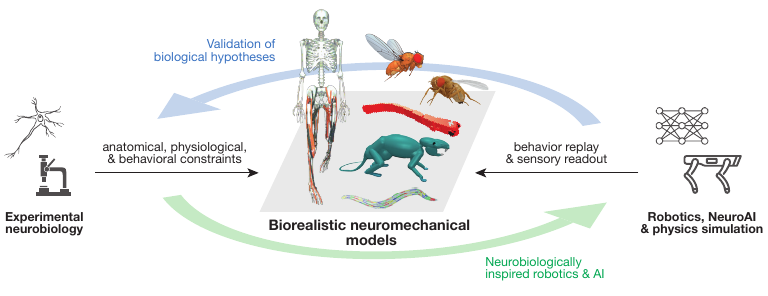}
    \caption*{%
        \textbf{Image attributions:}
        human model: MyoSuite \cite{vittorio_myosuite_2022};
        \textit{C. elegans} model: BAAIWorm \cite{zhao_baaiworm_2024};
        rodent model: MIMIC \cite{aldarondo_virtual_2024};
        zebrafish model: simZFish \cite{liu_artificial_2025};
        walking \textit{Drosophila} model: NeuroMechFly \cite{lobato-rios_neuromechfly_2022,wang-chen_neuromechfly_2024};
        flying \textit{Drosophila} model: FlyBody \cite{vaxenburg_whole-body_2025}.%
    }
    \label{fig:graphical_abstract}
\end{figure}

\vspace{0.5em}
\subsection*{Highlights}
\begin{itemize}
    \itemsep0em
    \item Biorealistic neuromechanical models comprise a neural controller and a biomechanical hull
    \item Replaying experimental recordings \textit{in silico} enables readout of latent variables
    \item Simulations help to test neuroscientific hypotheses
    \item Simulations provide a fully accessible and perturbable surrogate for behavior control
    \item Embodied computational models provide a foundation for NeuroAI
\end{itemize}

\vspace{\fill}
\begin{center}
    \footnotesize
    \copyright\ 2026. This manuscript version is made available under the CC-BY 4.0 license.\\
    \url{https://creativecommons.org/licenses/by/4.0/}
\end{center}

\clearpage
\newpage
\section*{Abstract} 
Animal behavior reflects interactions between the nervous system, body, and environment. Therefore, biomechanics and environmental context must be considered to understand algorithms for behavioral control. Computational models that embed artificial neural controllers within body models in simulated environments are a powerful tool for this purpose. Here, we review advances in biorealistic neuromechanical models while also highlighting emerging opportunities ahead. We first show how these models enable inference of biophysical variables that are difficult to measure experimentally. Through systematic perturbations, one can generate new experimentally testable hypotheses using these models. We then examine how neuromechanical models facilitate the exchange among neuroscience, robotics, and machine learning, and showcase their applications in healthcare. We envision that coupling experimental studies with active probing of their neuromechanical surrogates will significantly accelerate progress in neuroscience.

\vspace{1em}%
\noindent%
\textbf{Keywords:} Neuromechanical model, biomechanics, biorealism, artificial neural network, sensorimotor control, physics simulation, NeuroAI

\vspace{2em}
\section*{Introduction}

Animal behaviors cannot be understood in isolation from the body and the environments in which they are expressed \cite{krakauer_neuroscience_2017} because they arise from two tightly interconnected feedback loops: one between the nervous system and the body and the other between the body and external environment. Studying any one component alone, although valuable for addressing specific questions, can only provide a partial view of the integrative whole.

A wide range of experimental tools have been developed to investigate neural circuits (\textit{e.g.}, optogenetics, optical recordings, electrophysiology), measure body movements (\textit{e.g.}, motion capture, machine learning--based pose estimation), and manipulate environments (\textit{e.g.}, virtual reality). However, these tools face two fundamental limitations. First, there are considerable \textit{practical barriers} to obtaining neuronal recordings. For example, in large animals like primates, genetically targeting specific neurons remains challenging, and high-density electrode arrays cannot yet resolve individual cell identities. Additionally, muscle physiology measurements can interfere with natural behavior, limiting the number of muscles that can be recorded simultaneously without significantly hindering behaviors. Second, establishing causality in neuroscience typically require \textit{perturbations}: for example, artificial activation or silencing is typically used to support statements such as ``the firing of neuron A causes behavior B.'' However, most existing neural perturbation approaches are either too local (\textit{e.g.}, optogenetic manipulation of small numbers of neurons) or too broad (\textit{e.g.}, global pharmacological or genetic perturbations based on brain regions or cell types). Testing hypotheses about network-level dynamics would require patterned manipulation of large numbers of neurons with high specificity, but such experimental capabilities remain mostly out of reach.

\begin{figure}[p]
    \centering
    \hspace*{-0.13\textwidth}
    \includegraphics[width=1.2\linewidth]{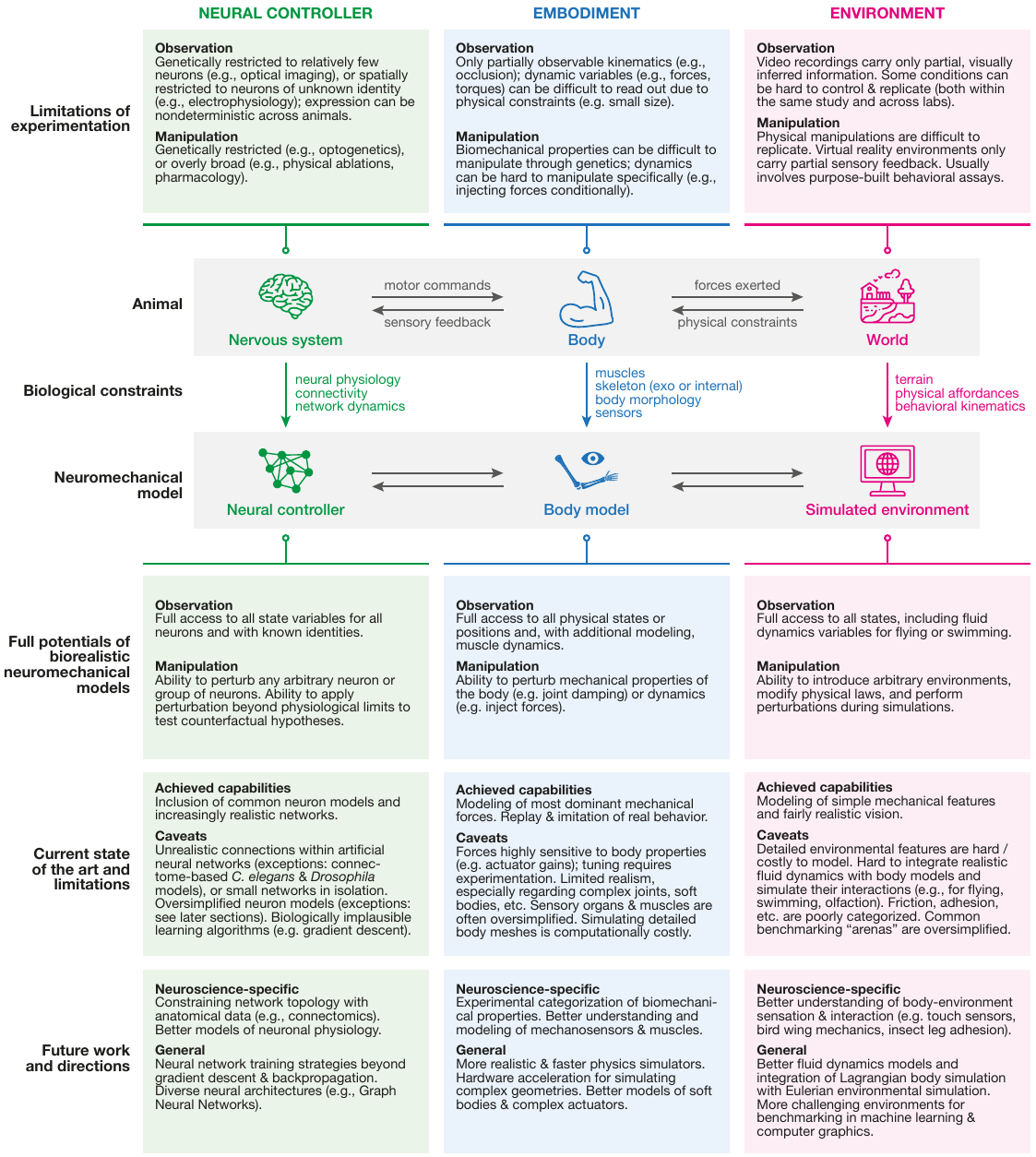}
    \caption{\textbf{Parallels between real animals and their biorealistic neuromechanical models.} Comparison of how physiological, mechanical, and environmental variables can be accessed and manipulated in real animals \textbf{(upper half)}, or in neuromechanical models \textbf{(lower half)}. Rows on the lower half summarize the full potential of neuromechanical models, the current status of the field, and key challenges that need to be overcome to reach this full potential. Relevant experimental ground truth data that can constrain neuromechanical modeling are indicated.}
    \label{fig:animal_vs_model}
\end{figure}

Models provide a complementary framework to address these limitations in experimental measurement and perturbation. These models range in complexity as follows:
\begin{itemize}
    \item \textbf{Biomechanical/body model:} A morphologically biorealistic representation of an animal (or its subparts) with \textit{articulated joints}, physically meaningful geometry, mass, and contact dynamics with the environment.
    \item \textbf{\textit{Neuro}mechanical model:} A biomechanical model that is coupled with emulation of \textit{neural dynamics}. If one aims to integrate biological actuators (e.g., motor neurons driving muscles and tendons) and sensors (e.g., muscle spindles and tactile sensors feeding back to inform the motor system), the biomechanical model must have some degree of morphological realism.
    \item \textbf{Neuromechanical digital twin:} A biorealistic neuromechanical model of a particular animal that is embedded in a physics simulator and updated continuously using real experimental measurements, ideally in real-time \cite{grieves2023digital}.
\end{itemize}
In neuromechanical models, at least in principle, body and neural controller models form a fully accessible closed-loop system in which sensorimotor dynamics can be generated, observed, and manipulated \textbf{(\autoref{fig:animal_vs_model})}. Because these models exist entirely \textit{in silico}, they can be subject to arbitrary perturbations and counterfactual simulations, allowing researchers to probe causality through ``what if'' scenarios. Thus, neuromechanical models complement \textit{in vivo} experiments by bridging multiple levels of biological organization. The use of simulated bodies controlled by artificial neural networks dates back decades, most prominently in studies of stick insect and cockroach locomotion \cite{ayali_comparative_2015}. More recently, advances in simulation fidelity and scalability, including differentiable physics engines (\textit{e.g.}, Brax \cite{brax}), differentiable neuron modeling frameworks (\textit{e.g.}, Jaxley \cite{deistler_jaxley_2025}), and hardware-accelerated simulators using graphics processing units (GPUs, \textit{e.g.}, MuJoCo Warp \cite{todorov_mujoco_2012,mjwarp}) have enabled the simulation of increasingly realistic neuromechanical models.

\begin{figure}[!b]
    \centering
    \includegraphics[width=\linewidth]{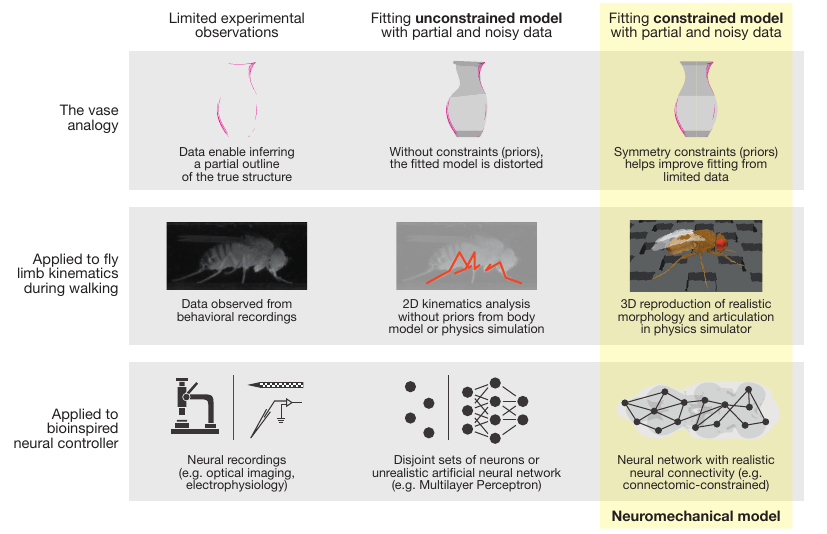}
    \caption{\textbf{Biological priors improve inference from limited experimental data.} \textbf{(Top row)} As an analogy, reconstructing a vase based on a partial sketch is under-constrained, and naive inference of the shape is prone to error. However, imposing a prior on vases (\textit{e.g.}, radial symmetry) helps infer missing information and consolidate redundant measurements. \textbf{(Middle and bottom rows)} Similarly, biologically grounded priors such as body morphology, musculoskeletal organization, and neural network topology constrain the solution space and improve reconstruction quality from sparse or noisy data. Because simulated dynamics are fully accessible, these models also enable inference of information beyond what is directly measurable experimentally.}
    \label{fig:vase_analogy}
\end{figure}

State-of-the-art neuromechanical models are constrained by biomechanical features. For example, joint articulations (\textit{e.g.}, degrees of freedom per joint and their ranges of motion) are derived from anatomy and joint movements are subject to passive dynamics such as stiffness and damping. Likewise, known neural circuit architecture (\textit{e.g.}, from connectomics) and neuronal properties (\textit{e.g.}, from electrophysiology) can be used to constrain the controller. Embedding such priors enables more effective and efficient use of limited experimental data. Akin to observing the shape of a vase from its partial outline yet being able to infer its complete shape using prior knowledge about vases (\textit{e.g.}, their symmetry), biologically grounded models improve inference when data are sparse or incomplete \textbf{(\autoref{fig:vase_analogy})}.

Despite these advantages, significant challenges remain. Artificial neural networks often lack biological realism in connectivity or physiological properties, while body models are often simplified using rigid bodies and lack detailed sensory organs and muscle models. Important biomechanical properties, such as actuator gains, are often left as free parameters due to the difficulties in measuring these values experimentally. Complex environments remain difficult to model, especially those involving fluid interactions for flight and swimming. Finally, even when more realistic simulation is possible, practical trade-offs exist between realism, computational cost, and implementation complexity. These challenges, along with current approaches to address them and work required going forward, are summarized in \textbf{\autoref{fig:animal_vs_model}} and discussed in the next sections.

In this review, we survey recent advances in biorealistic neuromechanical models and their applications to neuroscience, focusing primarily on work from the past five years while including widely used simulation tools and foundational studies. Our aim is to highlight opportunities for fundamental neuroscience and its dialogue with robotics and artificial systems. A summary of reviewed work is provided in \textbf{\autoref{tab:summary}}. We organize our perspective around five themes: (1) inferring unmeasurable variables from experimental data, (2) hypothesis testing by simulating forward-engineered models, (3) studying closed-loop behavior, (4) synergies with robotics and NeuroAI, and (5) healthcare applications.

\begin{table}[p]
    \centering
    \makebox[\textwidth][c]{%
        \newcommand{\outstandinginterest}{\textbullet\textbullet}
\newcommand{\specialinterest}{\textbullet}

{\scriptsize
\begin{tabular}{r@{\hspace{2pt}}llcll}
\toprule
 & \textbf{Article} &
     \textbf{Model organism} &
     \textbf{Year} &
     \textbf{Key component} &
     \textbf{Highlight} \\
     \midrule
\multicolumn{6}{l}{\textbf{Basic neuroscience}} \\
    \hdashline
\specialinterest
& Özdil \textit{et al.} \cite{ozdil_musculoskeletal_2025} &
    \textit{Drosophila} &
    2025 &
    Muscle actuation &
    Identify muscle synergies by motion replay \& imitation \\ 
    \hdashline
& Karashchuk \textit{et al.} \cite{karashchuk_sensorimotor_2024} &
    \textit{Drosophila} &
    2025 &
    Feedback loops &
    Modeling sensorimotor delay in locomotion \\ 
    \hdashline
\specialinterest
& Özdil \textit{et al.} \cite{ozdil_centralized_2026} &
    \textit{Drosophila} &
    2025 &
    Neural circuits &
    Probing grooming control with network \& body models \\ 
    \hdashline
\outstandinginterest
& Vaxenburg \textit{et al.} \cite{vaxenburg_whole-body_2025} &
    \textit{Drosophila} &
    2025 &
    Integrated &
    FlyBody: whole-body fly model and imitation learning \\ 
    \hdashline
\outstandinginterest
& Wang-Chen \textit{et al.} \cite{wang-chen_neuromechfly_2024} &
    \textit{Drosophila} &
    2024 &
    Integrated &
    NeuroMechFly v2: complex behaviors \& controllers \\ 
    \hdashline
& Lobato-Rios \textit{et al.} \cite{lobato-rios_neuromechfly_2022} &
    \textit{Drosophila} &
    2022 &
    Integrated &
    NeuroMechFly: realistic fly body model \\ 
    \hdashline
& Jin \textit{et al.} \cite{jin_wholebrain_2026} &
    \textit{Drosophila} &
    2026 &
    Neural circuits &
    Behavior control using connectome-based controller \\
    \hdashline
& Shiu \textit{et al.} \cite{shiu_drosophila_2024} &
    \textit{Drosophila} &
    2024 &
    Neural circuits &
    Whole-brain connectome simulation \\
    \hdashline
& Lappalainen \textit{et al.} \cite{lappalainen_connectome-constrained_2024} &
    \textit{Drosophila} &
    2024 &
    Neural circuits &
    Connectome-based visual system simulation \\
& Goldsmith \textit{et al.} \cite{goldsmith_neurodynamic_2020} &
    \textit{Drosophila}/robot &
    2020 &
    Embodiment &
    Drosophibot: fly-inspired hexapod robot \\ 
    \hdashline
\outstandinginterest
& DeWolf \textit{et al.} \cite{dewolf_neuro-musculoskeletal_2024} &
    Rodent (mouse) &
    2024 &
    Muscles &
    Muscle-level adaptive learning with MusBioMaus \\ 
    \hdashline
\outstandinginterest
& Aldarondo \textit{et al.} \cite{aldarondo_virtual_2024} &
    Rodent (rat) &
    2024 &
    Integrated &
    Realistic rodent model and imitation of real behavior \\ 
    \hdashline
\specialinterest
& Ravel \textit{et al.} \cite{ravel_modeling_2025} &
    Zebrafish &
    2025 &
    Environment &
    Escape swim with various water viscosities \\ 
    \hdashline
\specialinterest
& Liu \textit{et al.} \cite{liu_artificial_2025} &
    Zebrafish &
    2024 &
    Feedback loops &
    Zebrafish visuomotor behavior \\ 
    \hdashline
& Roussel \textit{et al.} \cite{roussel_modeling_2021} &
    Zebrafish &
    2021 &
    Neural circuits &
    Circuits driving movements in developing zebrafish \\ 
    \hdashline
\specialinterest
& Zhao \textit{et al.} \cite{zhao_roles_2025} &
    \textit{C. elegans} &
    2025 &
    Feedback loops &
    Feedback loops in \textit{C. elegans} locomotion \\ 
    \hdashline
\outstandinginterest
& Zhao \textit{et al.} \cite{zhao_baaiworm_2024} &
    \textit{C. elegans} &
    2024 &
    Integrated &
    BAAIWorm: simulate brain-body-environment loops \\ 
    \hdashline
& Olivares \textit{et al.} \cite{olivares_neuromechanical_2021} &
    \textit{C. elegans} &
    2021 &
    Neural circuits &
    Locomotion driven by pattern generator networks \\ 
    \hdashline
\specialinterest
& Pazzaglia \textit{et al.} \cite{pazzaglia_balancing_2025} &
    Salamander &
    2024 &
    Neural circuits &
    Balancing central control and sensory feedback \\ 
    \hdashline
& Thandiackal \textit{et al.} \cite{thandiackal_emergence_2021} &
    Lamprey/robot &
    2021 &
    Feedback loops &
    Swimming control via local force sensing \\ 
    \hdashline
& Pallasdies \textit{et al.} \cite{pallasdies_neuronal_2025} &
    Jellyfish &
    2025 &
    Neural circuits &
    Simulation of neuronal synchronization in swimming \\ 
    \hdashline
& Hoover \textit{et al.} \cite{hoover_neuromechanical_2021} &
    Jellyfish &
    2021 &
    Embodiment &
    Neuromechanical resonance benefits swimming \\ 
    \hdashline
\specialinterest
& Marin Vargas \textit{et al.} \cite{marinvargas_task-driven_2024} &
    Primate &
    2024 &
    Mechanosensation &
    Replay motion \textit{in silico} to test role of proprioception \\ 
    \hdashline
& Di Russo \textit{et al.} \cite{dirusso_sensory_2021} &
    Human &
    2021 &
    Feedback loops &
    Modeling gait patterns in human locomotion \\ 
    \hdashline
& Bellegarda \& Ijspeert \cite{bellegarda_cpg-rl_2022} &
    Robot &
    2022 &
    Neural circuits &
    Controlling pattern generators with deep learning \\ 
    \hdashline
& Stark \textit{et al.} \cite{stark_three-dimensional_2021} &
    Dog/robot &
    2021 &
    Embodiment &
    Dog musculoskeletal model \\
    \hdashline
& Housley \textit{et al.} \cite{housley_spindle_2023} &
    \textit{Nonspecific} &
    2023 &
    Mechanosensation &
    Biophysical model of proprioceptive encoding \\
    \hdashline
& Chacon \textit{et al.} \cite{chacon_spindle_2023} &
    \textit{Nonspecific} &
    2023 &
    Mechanosensation &
    Biophysical model of proprioceptive encoding \\
    \hdashline
& Sandbrink \textit{et al.} \cite{sandbrink_contrasting_2023} &
    Human &
    2023 &
    Mechanosensation &
    Machine learning model of proprioceptive encoding \\
    \hdashline
& Wang \textit{et al.} \cite{wang_octopus_2025} &
    Octopus &
    2025 &
    Mechanosensation &
    Mechanosensing in soft bodies \\
    \hdashline
& Mamiya \textit{et al.} \cite{mamiya_biomechanical_2023} &
    \textit{Drosophila} &
    2023 &
    Mechanosensation &
    High-res. neuroanatomy of sensory organs and neurons \\
    \hdashline
& Strauß \textit{et al.} \cite{strauss_neuroanatomy_2026} &
    \textit{Bacillus rossius} &
    2026 &
    Mechanosensation &
    High-res. neuroanatomy of sensory organs and neurons \\
    \midrule
\multicolumn{6}{l}{\textbf{Healthcare applications}} \\
    \hdashline
& Elisha \textit{et al.} \cite{elisha_esophagus_2025} &
    Human &
    2025 &
    Pathology &
    Esophagus biomechanical model and dysfunction \\ 
    \hdashline
& Singh \textit{et al.} \cite{singh_in-silico_2025} &
    Human &
    2025 &
    Pathology &
    Neuromechanical model for Spasticity progression \\ 
    \hdashline
& Sartori \textit{et al.} \cite{sartori_ceinms-rt_nodate} &
    Human/robot &
    2024 &
    Assistive devices &
    CEINMS-RT: individualized models \& wearable robots \\ 
    \hdashline
& Zhao \textit{et al.} \cite{zhao_neuromechanics-based_2023} &
    Human &
    2023 &
    Pathology &
    Model-based feedback controller for rehabilitation \\ 
    \hdashline
& Bruel \textit{et al.} \cite{bruel_investigation_2022} &
    Human &
    2022 &
    Pathology &
    Link biomechanical impairments to pathological gaits \\ 
    \hdashline
& Zhang \textit{et al.} \cite{zhang_ankle_2021} &
    Human &
    2021 &
    Algorithm &
    EMG + body model for ankle torque estimation \\
    \hdashline
& Meszaros-Beller \cite{meszarosbeller_individualisation_2023} &
    Human &
    2023 &
    Embodiment &
    Load distribution in individualized body models \\
    \midrule
\multicolumn{6}{l}{\textbf{Technological development}} \\
    \hdashline
& Todorov \textit{et al.} \cite{todorov_mujoco_2012} &
    \textit{Nonspecific} &
    2012 &
    Software &
    MuJoCo: a widely used physics simulator \\ 
    \hdashline
& MuJoCo developers \cite{mjwarp} &
    \textit{Nonspecific} &
    2025 &
    Software &
    MuJoCo Warp: GPU-accelerated physics simulation \\ 
    \hdashline
& Freeman \textit{et al.} \cite{brax} &
    \textit{Nonspecific} &
    2021 &
    Software &
    Brax: differentiable physics simulation \\ 
    \hdashline
& Deistler \textit{et al.} \cite{deistler_jaxley_2025} &
    \textit{Nonspecific} &
    2025 &
    Software &
    Jaxley: differentiable neuronal model simulation \\ 
    \hdashline
& Cotton \cite{cotton_kintwin_2025} &
    Human &
    2025 &
    Algorithm &
    Control motion by imitation at torque/muscle level \\ 
    \hdashline
& Vittorio \textit{et al.} \cite{vittorio_myosuite_2022} &
    Human &
    2022 &
    Software &
    MyoSuite: human-specific simulation environment \\ 
    \hdashline
& Feldotto \textit{et al.} \cite{feldotto_deploying_2022} &
    \textit{Nonspecific} &
    2022 &
    Software &
    Simulating embodied spiking networks at scale \\ 
    \hdashline
& Bohez \textit{et al.} \cite{bohez_imitate_2022} &
    Human &
    2022 &
    Algorithm &
    Modulate and reuse motor skills learned by imitation \\ 
    \hdashline
& Peng \textit{et al.} \cite{peng_learning_2020} &
    Dog/robot &
    2020 &
    Algorithm &
    Learning robotic control by imitating animals \\ 
    \hdashline
& Zhang \textit{et al.} \cite{zhang_mimicmjx_2025} &
    \textit{Nonspecific} &
    2025 &
    Software &
    Imitation learning through GPU-accelerated simulation \\ 
    \hdashline
& Merel \textit{et al.} \cite{merel_deep_2020} &
    Rodent &
    2020 &
    Algorithm &
    Interfacing biomechanical model with modern RL \\ 
    \hdashline
& Delp \textit{et al.} \cite{delp_opensim_2007} &
    \textit{Nonspecific} &
    2007 &
    Software &
    OpenSim: biomechanics-specific physics simulator \\ 
    \hdashline
& Yeo \textit{et al.} \cite{yeo_muscle_2023} &
    \textit{Nonspecific} &
    2023 &
    Muscles &
    Numerically stable muscle models \\
    \hdashline
& Keller \textit{et al.} \cite{keller_intrinsic_2025} &
    \textit{Zebrafish} &
    2025 &
    Algorithm &
    Ethologically inspired learning rule for artificial agents \\
& Wang-Chen \textit{et al.} \cite{wangchen_kinematics_2026} &
    \textit{Drosophila} &
    2026 &
    Experimentation &
    Synthetic data for experimental data analysis \\
    \midrule
\multicolumn{6}{l}{\textbf{Review \& perspective articles}} \\
    \hdashline
& Krakauer \textit{et al.} \cite{krakauer_neuroscience_2017} &
    \textit{Nonspecific} &
    2017 &
    Integrated &
    Importance of behavior in neuroscience studies \\
    \hdashline
& Bonanno \textit{et al.} \cite{bonanno_neural_2025} &
    Human &
    2025 &
    Pathology &
    Motor assessment of neurological disorders \\ 
    \hdashline
\specialinterest
& Zador \textit{et al.} \cite{zador_catalyzing_2023} &
    \textit{Nonspecific} &
    2023 &
    NeuroAI &
    Next-generation AI through neuroscience inspiration \\ 
    \hdashline
\outstandinginterest
& Ramdya \& Ijspeert \cite{ramdya_neuromechanics_2023} &
    \textit{Nonspecific} &
    2023 &
    NeuroAI &
    Synergy between neuroscience and robotics \\ 
    \hdashline
& Saxe \textit{et al.} \cite{saxe_if_2020} &
    \textit{Nonspecific} &
    2020 &
    NeuroAI &
    How deep learning helps understand the brain \\
    \hdashline
& Taiar \textit{et al.} \cite{taiar_editorial_2022} &
    Human &
    2022 &
    Pathology &
    Clinical link between circuits and biomechanics \\ 
    \hdashline
& Rodrigues da Silva \textit{et al.} \cite{rodriguesdasilva_comprehensive_2022} &
    Human &
    2022 &
    Assistive devices &
    Body models for device-assisted locomotion \\ 
    \hdashline
& De Groote \& Falisse \cite{de_groote_perspective_2021} &
    Human &
    2021 &
    Integrated &
    Predictive simulations of human movement \\ 
    \hdashline
& Merel \textit{et al.} \cite{merel_hierarchical_2019} &
    \textit{Nonspecific} &
    2019 &
    NeuroAI &
    Hierarchical motor control in animals and AI \\ 
    \hdashline
& Ayali \textit{et al.} \cite{ayali_comparative_2015} &
    Insects &
    2015 &
    Integrated &
    Foundational works on insect biomechanical models \\ 
    \hdashline
& Ostrow \textit{et al.} \cite{ostrow_comparing_2025} &
    \textit{Nonspecific} &
    2025 &
    Algorithm &
    Comparing network-level neural dynamics \\
    \hdashline
& Chung \textit{et al.} \cite{chung_geometry_2021} &
    \textit{Nonspecific} &
    2021 &
    Algorithm &
    Comparing network-level neural dynamics \\
    \hdashline
& Draelos \textit{et al.} \cite{draelos_improv_2025} &
    \textit{Nonspecific} &
    2021 &
    Experimentation &
    Closed-loop experimentation \\
    \bottomrule
\end{tabular}
}
    }
    \caption{\textbf{Overview of reviewed works.} Articles of special interest or outstanding interest (denoted with {\textbullet} or {\textbullet\textbullet} respectively) are highlighted in \hyperref[sec:annotated_refs]{Annotated References}.}
    \label{tab:summary}
\end{table}

\section*{Inference of experimentally unmeasurable variables}

\begin{figure}[t]
    \centering
    \hspace*{-0.13\textwidth}
    \includegraphics[width=1.2\linewidth]{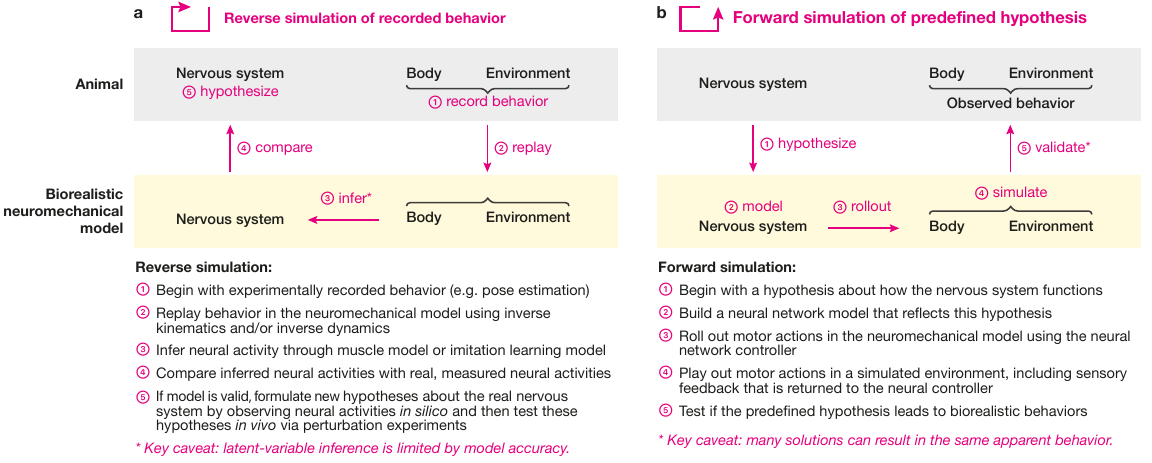}
    \caption{\textbf{Two directions for linking experiments to neuromechanical simulations.} (\textbf{a}) \textit{Reverse simulations:} replaying experimental recordings \textit{in silico} to infer hidden variables (\textit{e.g.}, neural and muscle activities) that are hard to measure directly. (\textbf{b}) \textit{Forward simulations:} generating behaviors from forward-engineered, hypothesis-driven models and comparing them with experimental observations.}
    \label{fig:reverse_vs_forward}
\end{figure}

One obstacle to understanding how the brain controls behavior is that many variables in sensorimotor control are difficult to measure experimentally. For example, internal forces are hard to measure non-invasively, muscle activities are difficult to record at scale, and sensory feedback is often only indirectly inferred. Biorealistic neuromechanical models can help infer these variables using only partial experimental data \textbf{(\autoref{fig:reverse_vs_forward}a)}. Typically, this begins with \textit{inverse kinematics}---estimating joint angles or body configurations from recorded movements---followed by \textit{inverse dynamics}---computing forces required to generate such movements. Inferred quantities include internal forces \cite{lobato-rios_neuromechfly_2022,ozdil_centralized_2026,zhang_ankle_2021}, muscle activities \cite{ozdil_musculoskeletal_2025,dewolf_neuro-musculoskeletal_2024,dirusso_sensory_2021,stark_three-dimensional_2021,bruel_investigation_2022}, sensory feedback \cite{wang-chen_neuromechfly_2024,marinvargas_task-driven_2024,dirusso_sensory_2021}, and energy consumption \cite{ravel_modeling_2025}. These inferred quantities provide a more direct link from kinematics to force generation (through muscles) and sensory experience. They also enable new ways of quantifying behavior, for example by using contact forces as a proxy for grooming vigor \cite{ozdil_centralized_2026}. Finally, a simpler use of body models is to generate synthetic behavior recordings paired with full kinematic state information, supporting the training of machine learning models for automated behavior analysis \cite{wangchen_kinematics_2026}.

Replaying experimentally recorded behavior also enables inference of \textit{neuromuscular dynamics}. This inference is inherently challenging because of muscle redundancy (more muscles than joint degrees of freedom), meaning that many muscle recruitment patterns can produce the same movement. One approach to combat this challenge is to impose criteria that favor some solutions over others: for example by minimizing total muscle activation or forces at any snapshot (\textit{e.g.}, OpenSim's static optimization \cite{delp_opensim_2007}). This approach has already provided insights: for instance, by identifying muscle co-activation patterns and suggesting potential muscle synergies \cite{ozdil_musculoskeletal_2025,stark_three-dimensional_2021}. A more dynamically constrained approach is to train models that predict the muscle activities needed to replicate observed \textit{sequences} of movements over time \cite{ozdil_musculoskeletal_2025,cotton_kintwin_2025}. This is a form of \textit{imitation learning} (IL) that we will revisit below.

Estimated muscle dynamics serve as a gateway to probe the \textit{neural control} of behavior. For example, passively replaying experimentally recorded primate reaching behavior in a musculoskeletal model allowed Marin Vargas \textit{et al.} to evaluate which information is best encoded in spindle responses inferred from the replay, thereby testing hypothesized roles for proprioception \cite{marinvargas_task-driven_2024}. In another example, Aldarondo \textit{et al.} trained an artificial neural network to replicate recorded behavior---effectively training an inverse dynamics controller---and identified candidate brain regions for similar inverse dynamics computing by comparing activities in the simulated and the real animal \cite{aldarondo_virtual_2024}. Neuromechanical models have also been extended beyond mechanosensation: for instance, studies of visuomotor behaviors have directly accessed simulated visual system activities using artificial neural networks constrained by biological circuitry \cite{liu_artificial_2025,wang-chen_neuromechfly_2024}.

An important caveat of reverse simulation is that inferred variables depend heavily on model accuracy. Simulations can easily produce seemingly plausible results and this accessibility can create a false sense of confidence. It is critical that inferred results should be interpreted consistently with the modeler's confidence in the model, though the level of accuracy required depends on the specific task. Recently, studies have improved \textit{anatomical} realism using high-resolution body scans \cite{mamiya_biomechanical_2023,strauss_neuroanatomy_2026}; in parallel, \textit{functional} models of sensory systems, including in muscles (as proprioceptive sensors) \cite{housley_spindle_2023,chacon_spindle_2023,sandbrink_contrasting_2023} and in soft materials \cite{wang_octopus_2025}. Another key caveat is the distinction between parameters that serve purely to improve simulation realism and those that represent biological quantities. The former, such as simulation timestep and loss function, can be considered modeling artifacts and tuned freely. The latter, such as joint stiffness, muscle properties, and neural connections, should not be tuned simply to improve the simulation. While both are often fit using similar optimization methods, their fundamentally different interpretations should not be obscured. Deciding how to treat each parameter is therefore not merely technical, but a conceptual choice to be made explicitly by the modeler.

\section*{Testing hypothesized neural controllers through simulation}

In the preceding section, we reviewed how neuromechanical models can replicate observed behaviors and infer physiological variables \textbf{(\autoref{fig:reverse_vs_forward}a)}. Conversely, forward-engineered controllers can be embedded in body models, and their simulated behaviors can be compared with experimental data \textbf{(\autoref{fig:reverse_vs_forward}b)}. Such models have been widely used to study locomotion driven by centralized oscillators (central pattern generators, CPGs) \cite{liu_artificial_2025,lobato-rios_neuromechfly_2022,olivares_neuromechanical_2021,bellegarda_cpg-rl_2022}, decentralized controllers based on mechanosensory feedback \cite{ayali_comparative_2015,dirusso_sensory_2021}, and hybrid approaches \cite{wang-chen_neuromechfly_2024,thandiackal_emergence_2021}. These models enabled researchers to investigate how low-dimensional modulatory signals modify patterned behavior, for example to steer \cite{liu_artificial_2025,wang-chen_neuromechfly_2024,bellegarda_cpg-rl_2022} or adjust walking gaits (\textit{e.g.} step length, cadence) \cite{dirusso_sensory_2021,bellegarda_cpg-rl_2022}. Similarly, neuromechanical models have been used to generate experimentally testable hypotheses on how timing and synchronization in neural circuits regulate coordinated behavior \cite{karashchuk_sensorimotor_2024,pallasdies_neuronal_2025}.

A key strength of neuromechanical simulations is their ability to extend neural control of behavior to the biomechanical embodiment that implements these control signals in naturalistic settings, which are difficult to study experimentally because neural recordings typically impose body constraints through, for example, tethering. For instance, Hoover \textit{et al.} \cite{hoover_neuromechanical_2021} embedded a neuromechanical model within a fluid dynamics simulation to show that the propagation speed of muscle activation in the jellyfish bell matches elastic deformation of the body, producing a resonance that enables efficient swimming. Finally, neuromechanical models provide a platform to study development, learning, and adaptation in motor control. For instance, Roussel \textit{et al.} \cite{roussel_modeling_2021} modeled putative neural circuits across zebrafish developmental stages to replicate the emergence of distinct swimming patterns. In another example, DeWolf \textit{et al.} \cite{dewolf_neuro-musculoskeletal_2024} replicated mouse behavior in a body model to test how sensorimotor prediction errors might be computed in the brain during adaptation to perturbations.

\section*{Closing the sensorimotor loop}

In the real world, behavior arises from nested feedback loops. For example, animals navigate the environment using sensory feedback that is continuously updated during self-motion. For example, the nervous system receives feedback about ongoing movements through proprioception. Embodying neural circuit models within a biomechanical hull in a physics simulator closes this first feedback loop \cite{wang-chen_neuromechfly_2024,vaxenburg_whole-body_2025}. Placing these body models in sensory-rich environments enables exploration of a second feedback loop \cite{aldarondo_virtual_2024,karashchuk_sensorimotor_2024,dewolf_neuro-musculoskeletal_2024}. For example, researchers have simulated a fly model driven by a hierarchical controller that integrates visual and olfactory cues to perform both high-level navigation and low-level locomotor movements to avoid obstacles on uneven terrain \cite{wang-chen_neuromechfly_2024}. For another fly model, a phenomenological fluid mechanics simulation was used, allowing the model to fly through a valley-like corridor using visual information \cite{vaxenburg_whole-body_2025}. Similarly, a combined brain-and-body model in the worm \textit{C. elegans} has been used to simulate movements toward attractors using sensory feedback \cite{zhao_baaiworm_2024}, consistent with experimentally observed behavior. These studies highlight how coupled neural and biomechanical models can be used to simulate complex sensorimotor behaviors that resemble those observed in nature.

Simulations also enable systematic perturbations to explore the mechanisms underlying closed loop behaviors, such as feedback-based locomotor control. This capability helps overcome an absence of experimental approaches where, for example, genetic manipulations are limited by the availability and specificity of transgenic driver lines. By contrast, \textit{in silico} models allow unrestricted perturbation of both the neural controller and the body. For example, Thandiackal \textit{et al.} \cite{thandiackal_emergence_2021} leveraged this flexibility to study swimming in a lamprey-like model (as well as in a physical robot). They systematically evaluated the robustness of different hypothesized neural controllers against neuromechanical perturbations, including ethically constrained manipulations such as spinal transection.

Finally, recent studies have advanced toward using neural networks with increasing biological fidelity. This includes incorporating (i) connectome-constrained artificial neural architectures \cite{wang-chen_neuromechfly_2024,shiu_drosophila_2024,jin_wholebrain_2026}, (ii) experimentally constrained circuit mechanisms \cite{zhao_roles_2025}, and (iii) physiologically realistic neuron models \cite{pazzaglia_balancing_2025,shiu_drosophila_2024}. For example, Wang-Chen \textit{et al.} \cite{wang-chen_neuromechfly_2024} fit a neuromechanical model of the fly with a visual system model \cite{lappalainen_connectome-constrained_2024} based on the real fly's brain connectome, enabling the inference of biologically plausible visual system responses which were then used to regulate simulated fly--fly chasing. These approaches demonstrate the co-dependence between brain and body models, which is essential to faithfully integrate realistic neural, biomechanical, and environmental dynamics in simulations.

\section*{Synergies with robotics and NeuroAI}

Neuroscience and robotics have long shared a common interest: understanding how autonomous agents---biological or artificial---behave both flexibly and robustly in complex environments \cite{ramdya_neuromechanics_2023}. Studies of animal behaviors have inspired robotic design \cite{goldsmith_neurodynamic_2020,peng_learning_2020,keller_intrinsic_2025}, while robots have served as physical testbeds for investigating neural control and sensorimotor integration in animals \cite{thandiackal_emergence_2021,goldsmith_neurodynamic_2020}. More recently, rapid progress in machine learning has renewed interest in the dialogue between artificial and biological intelligence under the framework of ``NeuroAI,'' which encompasses both neuroscience-inspired AI \cite{zador_catalyzing_2023} and the use of AI systems to evaluate and refine neuroscientific hypotheses \cite{saxe_if_2020}. Neuromechanical models provide a natural bridge between these two domains: (i) machine learning algorithms can be embodied in biologically realistic body models to test if they can explain observed animal behaviors and (ii) principles of biological behavioral control can be validated in simulation and used to improve AI systems.

A prominent aspect of integration between neuroscience and machine learning has been the use of reinforcement learning to replicate naturalistic behavior in biomechanical models. In particular, imitation learning (IL) \cite{bohez_imitate_2022,peng_learning_2020}, where autonomous agents learn directly from behavior recordings (``demonstrations'' in IL terminology) rather than abstract reward signals, has been especially effective. Combined with advances in motion tracking and pose estimation, IL has enabled realistic replication of experimentally recorded behaviors in neuromechanical models across species \cite{aldarondo_virtual_2024,vaxenburg_whole-body_2025,ozdil_musculoskeletal_2025,cotton_kintwin_2025,zhang_mimicmjx_2025} and allowed researchers to address a variety of neuroscientific questions \cite{aldarondo_virtual_2024,ozdil_musculoskeletal_2025,dewolf_neuro-musculoskeletal_2024}.

Another point of convergence between neuroscience and artificial systems is the hierarchical organization of neural architecture. Biological motor control happens at different levels: high-level processes (\textit{e.g.}, sensory integration, memory, decision-making) generate lower-dimensional signals that are conveyed to drive downstream motor circuits, where further neural computations transform these representations into low-level motor neuron outputs. Artificial neural controllers can mirror this architecture \cite{merel_hierarchical_2019,merel_deep_2020}: for example, fly models comprising multiple levels of neural processing (\textit{i.e.}, visual system, action selection, low-level motor control) have been used to produce naturalistic behaviors \cite{wang-chen_neuromechfly_2024,vaxenburg_whole-body_2025}. Conversely, such models can inform neuroscience. For instance, Karashchuk \textit{et al.} \cite{karashchuk_sensorimotor_2024} used a three-stage artificial controller inspired by the anatomical architecture of \textit{Drosophila} motor circuits to investigate how fast, robust walking emerges in the face of significant sensorimotor delays.

Beyond these aspects, NeuroAI also encompasses broader principles of behavioral control and learning. For example, insights from animal ethology have informed the design of optimization objectives for artificial agents (\textit{e.g.}, by using a generalized, intrinsic motivation for learning rather than task-specific rewards \cite{keller_intrinsic_2025}). In parallel, neuromechanical models have inspired the physical design of robots. For instance, decentralized control in a lamprey-like model that enables robust swimming after spinal injury offers a potential solution for modular robots \cite{thandiackal_emergence_2021}, neuromechanical resonance in jellyfish swimming can motivate similar designs in soft-bodied robots \cite{hoover_neuromechanical_2021}, and \textit{Drosophila} limb anatomy has informed the design of a hexapod robot \cite{goldsmith_neurodynamic_2020}.

\section*{Healthcare applications}

In humans, the integration of biomechanical models and neurological data provides a powerful tool for neurorehabilitation \cite{zhao_neuromechanics-based_2023}, understanding diseases \cite{singh_in-silico_2025,bruel_investigation_2022}, developing assistive medical devices \cite{sartori_ceinms-rt_nodate,zhang_ankle_2021}, and studying human physiology in general \cite{de_groote_perspective_2021}. For example, Bruel \textit{et al.} \cite{bruel_investigation_2022} simulated a leg musculoskeletal model using reflex-based control, which allowed them to identify which neural or biomechanical impairments could lead to pathological toe and heel gaits observed in patients with spinal injury, stroke and cerebral palsy. Similarly, Elisha \textit{et al.} \cite{elisha_esophagus_2025} replicated pathological movements in an esophagus biomechanical model, which, through systematic neuromechanical perturbation, generated hypotheses on the mechanisms of esophageal motility disorders. Biorealistic neuromechanical models can also be used for individualized medicine where models are constrained by a particular patient's physiological recordings (\textit{e.g.}, computed tomography or electromyography data) \cite{sartori_ceinms-rt_nodate,zhang_ankle_2021}. These models reveal important inter-individual variabilities---such as differences in load sharing across muscles \cite{meszarosbeller_individualisation_2023}---that may be important for clinical treatment. When these models are then updated with new patient data, they can be referred to as `digital twins'. A series of review articles detail how neuromechanical models can be applied in medical and clinical settings \cite{taiar_editorial_2022}, particularly within the aim of developing medical devices \cite{rodriguesdasilva_comprehensive_2022} and treating disorders \cite{bonanno_neural_2025}.

\section*{Outlook}

Biorealistic neuromechanical models are emerging as an integrative tool to link neural circuits, biomechanics, and behavior. Concurrently, rapid advances in connectomics, large-scale physiological recordings, and high-performance computing are expanding both the data and tools available to constrain these models. In this context, we anticipate several promising directions for the future.

First, more realistic models will become available. Increasingly detailed whole-body scans and advances in soft-body simulation will enable more accurate representations of muscles, tendons, compliant tissues, and soft materials, improving both motor output and mechanosensory feedback. On the neural side, connectomics datasets will impose strong architectural constraints on simulated networks, while large-scale simulations of physiologically realistic neurons \cite{feldotto_deploying_2022,shiu_drosophila_2024}, informed by recordings of large neuronal populations, will help bridge network structure and function. Together, these advances will also facilitate individualized models that capture subject-specific morphology, physiology, and pathology \cite{meszarosbeller_individualisation_2023} for both basic neuroscience and clinical research.

Second, we expect neuromechanical models to become not just an interpretive tool, but also an active part of experimental pipelines. Real-time simulations informed by ongoing measurements will enable closed-loop experimentation \cite{draelos_improv_2025}, where variables inferred from the model can guide experimental manipulations and measures on the fly, analogous to how smart microscopy uses online image analysis to steer data acquisition. This approach will allow researchers to more efficiently investigate hard-to-access internal states, explore atypical but important behaviors, and possibly perturb the system to bring it outside its normal operating states to perform counterfactual experiments. In the more distant future, simulation-in-the-loop experimentation may shift the design of neuroscience studies from largely \textit{post hoc} analysis of acquired data (though the experiment itself might involve artificial perturbations) to active probing of brain-body-environment interactions.

Third, neuromechanical models offer an opportunity to identify and validate generalizable principles of sensorimotor control. Comparative studies across species may reveal recurring motifs in control architectures (\textit{e.g.}, hierarchical versus decentralized), sensorimotor strategies (\textit{e.g.}, feedforward versus feedback-based), mechanical strategies (\textit{e.g.}, passive dynamics, compliance), and scaling laws (\textit{e.g.}, across different body sizes and configurations). Achieving this will require studying broader and more naturalistic behaviors beyond stereotyped laboratory tasks and conventional model organisms. Resulting hypotheses can then be tested \textit{in silico} and possibly implemented in physical robots and other embodied AI systems \cite{zador_catalyzing_2023}.

Realizing this exciting future will require sustained interdisciplinary exchange and shared infrastructure, such as faster and more accurate simulation engines \cite{mjwarp,zhang_mimicmjx_2025}, scalable pipelines for body model construction, and interoperable software stacks. These needs reveal an underappreciated alignment across fields with vastly diverging goals but shared technical challenges, including computer graphics (efficient simulation and rendering), cinematic animation and gaming (realistic and controllable characters), sports science (individualized models), industrial digital twins (real-time updates), human-computer interaction and virtual/augmented reality (closed-loop experimentation), in addition to more conventional targets for collaboration such as machine learning (neural architectures and optimization) and control theory (system identification).

Achieving greater realism in neuromechanical models will depend on identifying further constraints on both the body and brain. This will require characterization of biomechanical properties (\textit{e.g.}, joint actuation, compliance, sensing) and neuronal properties (\textit{e.g.}, more diverse connectomics datasets, measurements of neuronal intrinsic properties). Combined neural and behavioral recordings at high spatiotemporal resolution will also be essential to constrain models using approaches to study high-dimensional network dynamics \cite{chung_geometry_2021,ostrow_comparing_2025}.

Parallel to building more realistic models, it is equally important to understand which level of detail is needed, and in which aspects of the model, to address different scientific questions. For example, accurate body geometry may be required to model contact-based behaviors like grooming but simplified body geometries, such as cylindrical leg segments, may be sufficient to reproduce key features of walking in a physics simulator. Similarly, detailed models of muscles, tendons, and mechanosensory organs may be necessary to investigate low-level sensorimotor circuits but not necessary to probe higher-level neural architectures for action selection and motor learning. Conversely, studying biomechanical features such as active grip control may require accurate end effector geometry and contact mechanics but not realistic neuronal physiology and connectivity. More generally, we hypothesize that \emph{a complete model capable of reenacting behavior end-to-end is beneficial even for studies of specific subparts of behavioral control because it enables hypothesis testing in a closed-loop, embodied setting. However, functions of other subparts can often be surrogated by simpler models to reduce unnecessary complexity and avoid over-parameterization.}

\enlargethispage{\baselineskip}  
Together, these developments will transform neuromechanical models from specialized tools into a general framework for studying the neural control of behavior, enabling hypothesis generation, guiding experiments, and uncovering general principles of sensorimotor control.

\clearpage
\newpage
\phantomsection
\section*{Annotated references for articles published since 2023}
\label{sec:annotated_refs}

\subsection*{Papers of outstanding interest (\textbullet\textbullet)}

\subsubsection*{Aldarondo \textit{et al.} (2024) \cite{aldarondo_virtual_2024}}
This study constructed a ``virtual rodent'' by embedding an artificial neural network within a biomechanically realistic rat model. The artificial neural controller was trained to imitate experimentally recorded behavior using reinforcement learning. The authors then showed that neural activity in the real animal’s striatum and motor cortex could be predicted by activity in the artificial neural network controlling the virtual rat’s movements. Finally, the authors showed evidence in favor of the minimal intervention principle of optimal feedback control by perturbing latent variables in the control network. 

\subsubsection*{DeWolf \textit{et al.} (2024) \cite{dewolf_neuro-musculoskeletal_2024}}
This study introduced a 50-muscle biomechanical model of the mouse forelimb, ``MusBioMouse.'' By replaying the recorded behavior \textit{in silico}, the authors inferred experimentally inaccessible control signals---namely joint torques and muscle activation---from the model's muscles. By correlating these with experimentally recorded neural activities, the authors showed that somatosensory and motor cortex encode postural, muscle, and proprioceptive information.

\subsubsection*{Ramdya and Ijspeert (2023) \cite{ramdya_neuromechanics_2023}}
This article reviews how the fields of neuroscience and robotics have fostered each other, centered around the shared goal of understanding how agile, efficient, and robust movements are controlled by autonomous agents. The article outlines the parallels between biological and artificial systems and surveys shared insights from the two fields of study in the past decades. 

\subsubsection*{Vaxenburg \textit{et al.} (2025) \cite{vaxenburg_whole-body_2025}}
This study introduced a \textit{Drosophila} full-body model. An artificial neural network trained through end-to-end reinforcement learning was used to control the body model to imitate experimentally recorded walking and flying behavior. The controller was hierarchical---allowing the simulated fly to steer and adjust flying altitude, thus navigating a terrain with bumps and trenches during flight based on visual input.  

\subsubsection*{Wang-Chen \textit{et al.} (2024) \cite{wang-chen_neuromechfly_2024}}
This study expanded the functionality of a biologically realistic fly body model, ``NeuroMechFly'' \cite{lobato-rios_neuromechfly_2022}, by introducing biologically inspired walking control networks and incorporating visual and olfactory inputs. Using these, the authors modeled closed-loop behaviors such as head stabilization while walking over rugged terrain and path integration. The study demonstrated the integrative nature of neuromechanical modeling by training a hierarchical controller to simulate olfactory and visual navigation while walking over rugged terrain. The study also simulated a connectome-constrained visual system model to infer activities of biological neurons during locomotion.

\subsubsection*{Zhao \textit{et al.} (2024) \cite{zhao_baaiworm_2024}}
This study simulated the three-way interaction between the nervous system, the body, and the environment in \textit{C. elegans}. Using this framework, the authors replicated zigzag movements toward attractors in the environment, which had been experimentally observed. Notably, the neuromechanical model incorporated realistic neuronal connections constrained by the connectome, and realistic muscle placements constrained by the body anatomy.

\vspace{2em}

\subsection*{Papers of special interest (\textbullet)}

\subsubsection*{Liu \textit{et al.} (2025) \cite{liu_artificial_2025}}
This study introduced a realistic neuromechanical model of the larval zebrafish, ``simZFish,'' to investigate the animal’s optomotor response. By systematically perturbing controller features \textit{in silico}---an impossible task in real animals—the authors demonstrated how embodiment is tightly coupled with neural networks in the control of behaviors. Finally, the study constructed a physical robot, ``ZFish,'' to validate these insights.

\subsubsection*{\"Ozdil \textit{et al.} (2025) \cite{ozdil_musculoskeletal_2025}}
This study identified muscle synergies in the \textit{Drosophila} forelimb based on neuromechanical simulation. To do so, the authors replayed experimentally recorded behavior in a neuromechanical model and inferred muscle activities via an inverse model. Then they revealed groups of muscles that are recruited in synergy by clustering muscle activities. The study also used imitation learning to construct an artificial neural controller that drives realistic behavior by actuating muscles.

\subsubsection*{\"Ozdil \textit{et al.} (2024) \cite{ozdil_centralized_2026}}
Animal behaviors often require the coordination of multiple body parts. This study used \textit{Drosophila} grooming as a model to study multi-body part coordination. The authors used neuromechanical simulations to infer contact forces between the limbs and the antennae, which are challenging to measure experimentally. The authors also used counterfactual simulations (replaying recorded behavior with different actuator gains) to understand the roles of different movements in achieving the behavioral goal.

\subsubsection*{Pazzaglia \textit{et al.} (2024) \cite{pazzaglia_balancing_2025}}
This study used a salamander neuromechanical model to understand the role of mechanosensory feedback—particularly stretch feedback—in swimming and walking. The study coupled the body mechanical model with a neural network model consisting of leaky integrate-and-fire (LIF) neurons to investigate how stretch feedback signals could help the nervous system generate and maintain locomotor patterns.

\subsubsection*{Ravel \textit{et al.} (2025) \cite{ravel_modeling_2025}}
Behaviors often emerge as a result of the interaction between the body and its environment. This study simulated the interaction between the body and the water surrounding it during zebrafish swimming. Using this approach, the study highlights mechanical strategies employed by the animal to efficiently propel itself in waters of different viscosities, especially in terms of power output. 

\subsubsection*{Marin Vargas \textit{et al.} (2024) \cite{marinvargas_task-driven_2024}}
This study replayed primate arm movements in a musculoskeletal model and read out muscle proprioceptive signals \textit{in silico}. By training machine learning models to perform tasks that reflect hypothesized roles of proprioception (\textit{e.g.,} recognizing stereotyped actions, estimating body kinematic states, and most efficiently encoding overall sensory information), the authors tested hypotheses on the roles played by proprioception.

\subsubsection*{Zador \textit{et al.} (2023) \cite{zador_catalyzing_2023}}
This opinion piece highlights how advances in NeuroAI will accelerate the development of next-generation, embodied artificial intelligence systems that function in the real world.

\subsubsection*{Zhao \textit{et al.} (2025) \cite{zhao_roles_2025}}
By simulating the locomotion of the worm \textit{C. elegans}, this study proposed a feedback-driven control mechanism that comprises motor neurons and muscles. Combined calcium imaging with neuromechanical simulation in its methodology, this study exemplifies how \textit{in vivo} observation and \textit{in silico} simulation can be synergistic, providing a powerful platform for neuroscientific discoveries.

\vspace{2em}
\section*{Declaration of generative AI and AI-assisted technologies in the writing process}
During the preparation of this work the authors used OpenAI ChatGPT to improve the readability and language of the manuscript. After using this tool/service, the authors reviewed and edited the content as needed and take full responsibility for the content of the published article.

\section*{Declaration of interest}
The authors declare no competing interests.

\section*{Acknowledgements}
SWC acknowledges support from a Boehringer Ingelheim Fonds PhD fellowship. PR acknowledges support from a Swiss National Science Foundation (SNSF) Project Grant (207806).

\section*{Author contributions}
\noindent \textbf{SWC} contributed to conceptualization, investigation, visualization, writing -- original draft, writing -- review and editing.

\noindent \textbf{PR} contributed to conceptualization, project administration, supervision, visualization, writing -- original draft, writing -- review and editing.

\vspace{2em}
\bibliography{ref.bib}

\end{document}